\documentclass[twocolumn,showpacs,superscriptaddress,nofootinbib]{revtex4-2}
\usepackage[latin9]{inputenc}
\usepackage{amsmath,bm}
\usepackage{bbold}
\usepackage{amssymb}
\usepackage{amsthm}
\usepackage{graphicx}
\usepackage{color}
\usepackage{stmaryrd}
\usepackage{hyperref}

\theoremstyle{plain}
\newtheorem*{theorem*}{Theorem}

\makeatletter

\@ifundefined{textcolor}{}
{%
 \definecolor{BLACK}{gray}{0}
 \definecolor{WHITE}{gray}{1}
 \definecolor{RED}{rgb}{1,0,0}
 \definecolor{GREEN}{rgb}{0,1,0}
 \definecolor{BLUE}{rgb}{0,0,1}
 \definecolor{CYAN}{cmyk}{1,0,0,0}
 \definecolor{MAGENTA}{cmyk}{0,1,0,0}
 \definecolor{YELLOW}{cmyk}{0,0,1,0}
}

\usepackage{color}
\usepackage{amsfonts}
\usepackage{graphics}
\usepackage{epsfig}\usepackage{amsthm}

\def\identity{\leavevmode\hbox{\small1\kern-3.8pt\normalsize1}}

\renewcommand{\epsilon}{\varepsilon}

\makeatother

\begin{document}

\title{Quantum neuromorphic approach to efficient sensing of gravity-induced entanglement}

\author{Tanjung Krisnanda}
\affiliation{School of Physical and Mathematical Sciences, Nanyang Technological University, 637371 Singapore, Singapore}

\author{Tomasz Paterek}
\affiliation{Institute of Theoretical Physics and Astrophysics, Faculty of Mathematics, Physics and Informatics, University of Gda\'{n}sk, 80-308 Gda\'{n}sk, Poland}

\author{Mauro Paternostro}
\affiliation{School of Mathematics and Physics, Queen's University, Belfast BT7 1NN, United Kingdom}

\author{Timothy C. H. Liew}
\affiliation{School of Physical and Mathematical Sciences, Nanyang Technological University, 637371 Singapore, Singapore}
\affiliation{MajuLab, International Joint Research Unit UMI 3654, CNRS, Universit\'{e} C\^{o}te d'Azur, Sorbonne Universit\'{e}, National University of Singapore, Nanyang Technological University, Singapore}

\begin{abstract}
The detection of entanglement provides a definitive proof of quantumness. Its ascertainment might be challenging for hot or macroscopic objects, where entanglement is typically weak, but nevertheless present.
Here we propose a platform for measuring entanglement by connecting the objects of interest to an uncontrolled quantum network, whose emission (readout) is trained to learn and sense the entanglement of the former.
First, we demonstrate the platform and its features with generic quantum systems.
As the network effectively learns to recognise quantum states, it is possible to sense the amount of entanglement after training with only non-entangled states.
Furthermore, by taking into account measurement errors, we demonstrate entanglement sensing with precision that scales beyond the standard quantum limit and outperforms measurements performed directly on the objects.
Finally, we utilise our platform for sensing gravity-induced entanglement between two masses and predict an improvement of two orders of magnitude in the precision of entanglement estimation compared to existing techniques.
\end{abstract}

\maketitle



\section{introduction}
Following the success of the use of neural networks across different fields of science~\cite{ching2018opportunities,topol2019high,hannun2019cardiologist,mehta2019high} for detecting patterns in data, proposals have been set forth in the quantum regime~\cite{fujii2017harnessing,qrpreview}.
In this direction, a particular quantum neural network architecture has emerged -- termed \emph{quantum reservoir processing}, in analogy to classical reservoir computing~\cite{montavon2012neural}. In such architecture, the quantum network serving as a processor is composed of randomly interacting quantum systems (the \emph{nodes}), not requiring precise control.
The function of this kind of network is learned in a training procedure that measures the system individually and fixes only a single output layer, which makes it experimentally friendly.
This architecture has been proposed for executing classical tasks~\cite{fujii2017harnessing,govia2021quantum,xu2021superpolynomial} (showing performance advantage over classical networks) and quantum tasks such as state characterisation~\cite{ghosh2019quantum,ghosh2020reconstructing}, quantum state preparation~\cite{ghosh2019quantum2,creating2020}, gate compression~\cite{ghosh2021realising}, and quantum metrology~\cite{krisnanda2022phase} (see Refs.~\cite{markovic2020quantum,qrpreview} for reviews).
Remarkably, for characterisation and metrological tasks, it is not necessary to perform correlation measurements, and it suffices to measure only local observables such as average occupation numbers or intensities of the network nodes.
The platform is versatile and it holds the potential to directly estimate important quantities such as quantum entanglement, which is the focus of our study. 

Entanglement is a special type of correlation between two or more objects, the presence of which witnesses their quantum nature~\cite{horodecki2009quantum}. 
In experiments involving objects that cannot be accessed directly, their quantum character could be revealed by using such inaccessible systems as mediators between two accessible probes.
The revelation of an entanglement gain between the probes then provides proof of another quantum signature -- known as quantum discord -- of the mediators~\cite{krisnanda2017revealing}.
This experimental scheme has been put forward as a proposal to probe quantum signatures of gravity through the observation of gravity-induced entanglement between masses~\cite{bose2017spin,marletto2017gravitationally,krisnanda2020observable} (see also Refs.~\cite{balushi2018,belenchia2018quantum,qvarfort2020,van2020quantum,rijavec2021decoherence,margalit2021realization,Pedernales2022} for recent developments and discussion).
This motivates the general framework presented in this paper, which is aimed at sensing (possibly weak) entanglement and its application to gravity-induced entanglement.

To date, there are essentially two main schemes proposed for the observation of gravity-induced entanglement, which suffer of different practical difficulties.
The Bose \emph{et al.}-Marletto-Vedral (BMV) scenario~\cite{bose2017spin,marletto2017gravitationally} requires preparation of a macroscopic superposition of each of two nearby massive bodies, whose later dynamics might showcase gravitational entanglement.
The challenging state-preparation stage is bypassed in the proposal of Ref.~\cite{krisnanda2020observable}, which resorts to continuous-variable (CV) entanglement between masses that begin in natural and easy-to-arrange Gaussian states. In this scheme, the entanglement detection remains as a demanding step. 
We show that a relatively simple neural-network architecture is sufficient to achieve a two-order-of-magnitude improvement in the precision with which entanglement of massive systems can be estimated, when compared to state-of-the-art values~\cite{palomaki2013entangling}.

Specifically, we utilise a reservoir quantum network (QN) for precise entanglement sensing.
In particular, quantum objects whose entanglement we want to scrutinise (the input) are put in contact with a QN. 
The observables from the QN -- which can be as simple as the mean excitation numbers of the nodes -- are post-processed through a single output layer.
This layer is trained so that the final output estimates quantum entanglement of the input objects.
Our general platform is particularly useful in situations where the input is not accessible for direct measurements, the latter are complicated (this is particularly the case for those that necessitate conditional or correlation measurements), or in cases where the input is less resilient to measurement errors than the QN.
First, we will introduce the general framework with generic quantum systems.
We show that a QN can learn from a random set of non-entangled input states and nevertheless is able to estimate the amount of  entanglement at the testing stage.
For a more realistic scenario considering measurement errors, we show that the entanglement precision scales better than 
$\Delta_E \propto 1/\sqrt{\mathcal{N}}$, where $\mathcal{N}$ is the number of measured observables.
We shall refer to $1/\sqrt{\mathcal{N}}$ scaling as the standard quantum limit (SQL).
Finally, we demonstrate an explicit application of our framework and its features to the recent endeavour whose goal is to reveal quantum features of gravity by measuring gravity-induced entanglement between masses. 
In particular, we show that measurements on cavity modes, which have interacted with the masses, can be post-processed to estimate the gravity-induced entanglement. Importantly, our approach offers better sensitivity compared to direct measurements on the masses.

\section{The general framework}
\begin{figure}[h]
\centering
\includegraphics[width=0.48\textwidth]{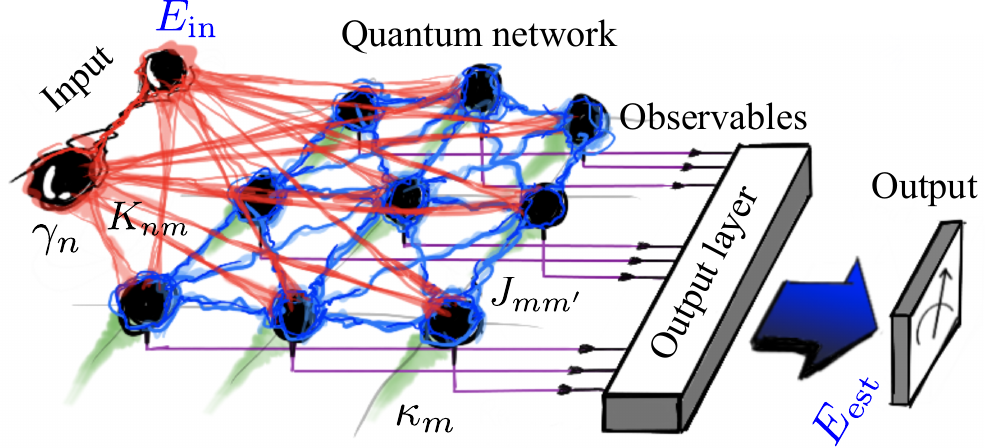}
\caption{Illustration of a quantum neuromorphic platform for entanglement sensing.
It involves to-be-measured input objects that are connected to a quantum network composed of uncontrolled nodes having random interactions with the input $K_{nm}$ and between themselves $J_{mm^{\prime}}$.
Both the input and the nodes interact with their environments, denoted by $\gamma_{n}$
and $\kappa_m$, respectively.
The observables from the QN are processed by a trained output layer, producing a signal that estimates input entanglement.
}
\label{Fig_framework}
\end{figure}

Our thought platform is depicted in Fig.~\ref{Fig_framework}.
Consider that quantum objects, whose entanglement is to be estimated, serve as the input. 
They come in contact with a processor, namely, a QN composed of quantum nodes.
Note that the QN nodes require minimal control, e.g., they can be randomly interacting with each other, the input, and environment.
We also allow that they are pumped by external coherent sources (e.g., in an optical system, lasers).
The purpose of the contact is a flow of information from the input to the QN. 
By retrieving the observables from the QN and processing them via a single output layer, one obtains a final output signal.
We will show that by training a set of weights and biases in the single output layer, the output signal estimates entanglement of the input objects.

Let us consider generic quantum systems and their dynamics with which we demonstrate the general framework described above.
In what follows, we consider continuous-variable systems (bosons).
Additionally, the platform also works for generic discrete systems (e.g., qubits), see \textcolor{blue}{Appendix~A} for details, as well as hybrid discrete-continuous systems, see \textcolor{blue}{Appendix~B}.
We begin by modelling the dynamics of the input $\rho_{\text{in}}$ and the QN $\rho_{\text{qn}}$ for a time $\tau$, after which the observables of the QN are recorded.
The coherent part of the dynamics is described by the following Hamiltonian, written in a frame rotating with the pump frequency
\begin{equation}
\begin{aligned}
    H&=\sum_{n = 1}^2 \hbar \Delta_n \hat a_n^{\dagger}\hat a_n + \sum_{m = 1}^M \hbar (\Lambda_m \hat b_m^{\dagger}\hat b_m+P_m (\hat b_m+\hat b_m^{\dagger}))  \\
    &+\sum_{\llbracket n,m\rrbracket} \hbar K_{nm} \mathcal{F}(\hat a_n,\hat b_m)+ \sum_{\llbracket m,m^{\prime}\rrbracket}\hbar J_{mm^{\prime}}\mathcal{F}(\hat b_m, \hat b_{m^{\prime}}),
    \label{EQ_gen_h}
\end{aligned}
\end{equation}
where $\hat a_n$ ($\hat b_m$) denotes the annihilation operator for the $n$th input object ($m$th QN node).
The detunings of all local frequencies $\{\omega_n,\Omega_m\}$ with respect to the frequency of the pump $\Theta_p$ are denoted by $\Delta_n=\omega_n-\Theta_p$ and $\Lambda_m=\Omega_m-\Theta_p$. 
The contact between the input and QN is represented by the couplings $K_{nm}$, whereas the interactions within the QN are denoted by $J_{mm^{\prime}}$. 
For simplicity, we take the operator function to represent interactions that are ample in nature, i.e., $\mathcal{F}(\hat X,\hat Y)\equiv \hat X \hat Y^{\dagger}+\hat Y \hat X^{\dagger}$.
Each QN node may be coherently driven with strength $P_m$.
The bracket $\llbracket \cdot,\cdot \rrbracket$ denotes a particular configuration of the couplings, e.g., all-to-all.


We note that the simulation of the system can be made efficient when dealing with Gaussian states~\cite{adesso2014continuous}.
These tools are applicable as the generic dynamics we consider here preserves Gaussianity, i.e., it involves a Hamiltonian that is at most quadratic in operators (Eq.~(\ref{EQ_gen_h})) and Gaussian dissipative processes (see below).
In this case, complete description of the system is contained in a covariance matrix (CM) $\bm V$ with elements $V_{ij}\equiv \langle u_i u_j +u_j u_i \rangle/2-\langle u_i \rangle \langle u_j \rangle$, where the vector ${\bm u}\equiv [\hat q_1,\hat r_1,\hat q_2,\hat r_2,\hat x_1,\hat p_1,\cdots,\hat x_M,\hat p_M]^T$ is composed of dimensionless position and momentum quadratures (of the input and QN nodes, respectively) that are expressed as $\hat q_n=(\hat a_n+\hat a_n^{\dagger})/\sqrt{2}$, $\hat r_n=(\hat a_n-\hat a_n^{\dagger})/(i\sqrt{2})$, $\hat x_m=(\hat b_m+\hat b_m^{\dagger})/\sqrt{2}$, and $\hat p_m=(\hat b_m-\hat b_m^{\dagger})/(i\sqrt{2})$.
One can obtain the dynamics of the quadratures in the Heisenberg picture from the Hamiltonian of Eq.~(\ref{EQ_gen_h}), which with added noise terms gives rise to a set of Langevin equations (LEs) that can be written in a matrix form: $\dot {\bm{u}}(t) = {\bm A} {\bm{u}}(t) + {\bm{h}}(t)$.
The drift matrix ${\bm A}$ contains the parameters $\{\Delta_n,\Lambda_m,K_{nm},J_{mm^{\prime}},\gamma_n,\kappa_m\}$ and the vector ${\bm h}(t)$ incorporates the pump and noise terms, see \textcolor{blue}{Appendix~C} for details.
The noise terms are of uncoloured Gaussian type, and written as $\sqrt{2\gamma_n}\hat a_n^{\text{in}}$ and $\sqrt{2\kappa_m}\hat b_m^{\text{in}}$, where $\langle \hat a_n^{\text{in}}(t)\hat a_{n^{\prime}}^{\text{in},\dagger}(t^{\prime})\rangle=\delta_{nn^{\prime}}\delta(t-t^{\prime})$ and $\langle \hat b_m^{\text{in}}(t)\hat b_{m^{\prime}}^{\text{in},\dagger}(t^{\prime})\rangle=\delta_{mm^{\prime}}\delta(t-t^{\prime})$~\cite{walls2007quantum}.

The solution of the LEs is given by 
\begin{equation}
{\bm u}(t)={\bm W}_+(t){\bm u}(0)+{\bm W}_+(t)\int_0^t dt^{\prime} {\bm W}_-(t^{\prime}){\bm h}(t^{\prime}),
\end{equation}
where ${\bm W}_{\pm}(t)=\exp{(\pm {\bm A}t)}$.
This further gives the dynamical equation for the CM: $\dot {\bm V}={\bm A}{\bm V}(t)+{\bm V}(t){\bm A}^T+{\bm D}$, where ${\bm D}=\mbox{diag}[\gamma_1,\gamma_1,\gamma_2,\gamma_2,\kappa_1,\kappa_1,\cdots,\kappa_M,\kappa_M]$.
The observables of the QN $\langle \hat O_{mk}\rangle=\mbox{tr}(\rho(\tau) \hat O_{mk})$ ($k$ labels observables from the same $m$th node) at time $\tau$ can be obtained from ${\bm u}(\tau)$ and ${\bm V}(\tau)$.
For more detailed expressions, see \textcolor{blue}{Appendix~C}.
Here we consider local observables, for simplicity.
We will see that it is sufficient to work with average occupation numbers (intensities) as the observables, although any additional variables that can be measured can further help.
The observables define an output layer upon which a training procedure is used to find a linear combination of the observables that will define the final system output.

The training is performed with ridge regression using a random set of input CMs $\{\bm V_{\text{in},l}\}_{l=1}^{N_{\text{tr}}}$ as follows.
Each of the input CMs will be in contact with the QN and produce a set of $N_{\text{ob}}$ observables at time $\tau$, recorded as a vector $\bm v_l$.
The observables are used to first estimate the input state (its \emph{unique} elements), from which entanglement is calculated.
In the present case, each element of the CM (labelled $s$) is estimated linearly as $f_s=\bm \beta_s  [1;\bm v_l]$, where $\bm \beta_s=[\beta_0,\beta_1,\cdots, \beta_{N_{\text{ob}}}]$ contains the coefficients to be obtained with ridge regression.
In particular, $\bm \beta_s=(\bm X^T\bm X+\lambda \mathbb{1})^{-1}\bm X^T\bm Y_s$, where $\bm X=[1,\bm v_1^T;1,\bm v_2^T;\cdots ;1,\bm v_{N_{\text{tr}}}^T]$ contains all the observables in the training set, $\bm Y_s$ contains the target $s$th element, and $\lambda$ is the ridge parameter.
This allows us to obtain an estimated input CM $\tilde {\bm V}_{\text{in}}$ from the trained output layer $\{\bm \beta_s\}$, given measured QN observables.
Consequently, the estimated entanglement is computed using the logarithmic negativity $E=L_e(\tilde {\bm V}_{\text{in}})$~\cite{negativity}.
In what follows, we define the entanglement estimation error as 
\begin{equation}\label{EQ_def_error}
\Delta_E=\sqrt{\sum_{l^{\prime} = 1}^{N_{\text{te}}} \frac{(E_{\text{est},l^{\prime}}-E_{\text{in},l^{\prime}})^2}{N_{\text{te}}}},
\end{equation}
where $N_{\text{te}}$ is the number of random input CMs in the testing set.

\section{Entanglement estimation}\label{S_ee}
Here we present the performance of entanglement estimation.
In simulations, the parameters are taken as random $\{\Delta_n,\Lambda_m,K_{nm},J_{mm^{\prime}},P_m,10\gamma_n,10\kappa_m\}\in [0,1]\Gamma$, where $\Gamma$ is an overall strength in units of frequency, and evolution time $\tau=\pi/2\Gamma$.
One set of random parameters will be taken to define one particular QN.
When assessing the performance of the scheme, we will average over different parameter choices, to provide a general assessment of the architecture rather than any specific parameter choice.
Indeed, one advantage of our scheme is that the considered systems do not need precise control of their parameters.

\begin{figure}[h]
\centering
\includegraphics[width=0.45\textwidth]{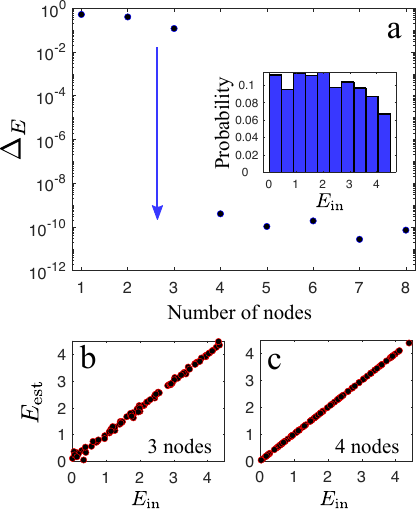}
\caption{Performance of entanglement sensing for generic dynamics.
(a) Estimation error vs number of nodes used in the QN.
The inset shows the profile of entanglement of the randomly generated input CMs in both training and testing. 
Panels (b) and (c) present explicit comparison between the estimated and input entanglement during testing where the QN is composed of $3$ and $4$ nodes, respectively.
}
\label{Fig2}
\end{figure}

Figure~\ref{Fig2}(a) shows the entanglement estimation error against the number of QN nodes. 
The sudden shift shown by the arrow indicates $\Delta_E\sim 10^{-10}$ is obtained for QNs having at least $4$ nodes.
This can be understood as follows.
Recall that the number of independent parameters required to fully characterise an $N$-mode Gaussian state is $2N(2N+1)/2$.
This suggests that to faithfully estimate the state of a two-mode Gaussian input, one requires at least $10$ observables from the QN. 
This is fulfilled by having at least $4$ QN nodes as each node is itself in a Gaussian state and hence requires three independent real parameters (e.g., we take two diagonal and one off-diagonal entries from the local CM) to be determined.
The inset shows the entanglement profile of the input CM ${\bm V}_{\text{in}}$ used in both training and testing, with $N_{\text{tr}}=50$ and $N_{\text{te}}=100$, respectively.
See \textcolor{blue}{Appendix~D} for the generation of random input CMs.
A closer look at the comparison between the estimated and input entanglement during testing is plotted in Figs.~\ref{Fig2}(b) and (c) for the case where the QN is composed of $3$ and $4$ nodes, respectively.
It can be seen that the latter offers minute errors.

We have also simulated the case where we record one observable (the mean excitation $\langle \hat b_m^{\dagger}\hat b_m \rangle$) from each QN node.
In this case, one requires either the addition of two-photon pump, i.e., $\sum_m P_m^{\prime}(\hat b^2_m+\hat b_m^{\dagger2})$ with random strengths (relatively weaker) $P_m^{\prime}\in [0,1]\Gamma/10$ or the presence of ultra-strong coupling $\mathcal{F}(\hat b_m,\hat b_{m^{\prime}})=(\hat b_m+\hat b_m^{\dagger})(\hat b_{m^{\prime}} +\hat b_{m^{\prime}}^{\dagger})$.
The reason for this is that simpler interactions or drives in Eq.~(\ref{EQ_gen_h}) are not sufficient for complex information transfer during the dynamics, which would allow the mean excitation of the QN nodes to completely recover information regarding the input objects. 
We found that the shift to low estimation error requires at least 10 QN nodes, again consistent with the number of independent parameters of the input CM, see \textcolor{blue}{Appendix~E} for details.



As the scheme estimates the CM of the input objects before computing entanglement, it opens up the possibility to use a training set consisting of separable input CMs without affecting its entanglement-testing capabilities.
We used the same setup as in Fig.~\ref{Fig2}(a) based on 4 QN nodes and performed training using only non-entangled input CMs.
The testing was performed with entangled input CMs, finding a profile similar to the inset in Fig.~\ref{Fig2}(a). 
Indeed, the comparison between the estimated and input entanglement is similar to the one in Fig.~\ref{Fig2}(c) (see \textcolor{blue}{Appendix~F} for details).
We note that although each input CM in the training set is not entangled, they are still correlated.
Similarly, learning from separable input objects is also possible for discrete systems (cf. \textcolor{blue}{Appendix~F}).

\section{Scaling beyond the SQL}
\begin{figure}[h]
\centering
\includegraphics[width=0.45\textwidth]{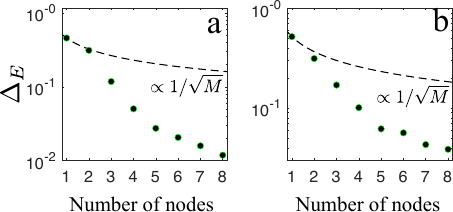}
\caption{Scaling beyond the SQL.
(a) Estimation error plotted against the number of nodes $M$.
Panel (b), taking only mean excitation from each QN node as the observable with time-multiplexing performed $\mathcal{T}=3$ times.
The scaling for the SQL is given by the dashed curve in each panel.
}
\label{Fig3}
\end{figure}

For a more realistic model, we incorporate measurement errors of the observables from the QN. 
The observables now read $\langle \hat O_{mk}\rangle \rightarrow \langle \hat O_{mk}\rangle+\epsilon_{mk}$, where $\{\epsilon_{mk}\}$ are generated from a normal distribution with zero mean and standard deviation $\zeta/2$.
In what follows, we take $\zeta=10^{-3}$. 
Similar to Fig.~\ref{Fig2}(a), we present the estimation errors in Fig.~\ref{Fig3}(a).
The dots indicate the scaling of the error with respect to the number of nodes $M$.
From Fig.~\ref{Fig3}(a), one can see the signature of the shift previously observed in Fig.~\ref{Fig2}(a). 
In particular, the scaling of the estimation error becomes clearer for $M\ge 4$ in Fig.~\ref{Fig3}(a).
It can be seen that $\Delta_E$ can exhibit scaling beyond the dashed curve, i.e., beyond the SQL $\propto 1/\sqrt{\mathcal{N}} = 1 / \sqrt{3M} \propto 1/\sqrt{M}$.

Another alternative to obtain independent observables from the QN is through time-multiplexing.
For instance, we consider a single observable from each QN node, i.e., the mean excitation $\langle \hat b_m^{\dagger}\hat b_m \rangle$ and measure it at $\mathcal{T}$ different times.
This gives a total of $N_{\text{ob}}=M\mathcal{T}$ observables.
We demonstrate the case for $\mathcal{T}=3$, i.e., at $\tau=\{1,2,3\} \pi/2\Gamma$ in Fig.~\ref{Fig3}(b). 
In this case, we have added random two-photon pump $P_m^{\prime}\in [0,1]\Gamma/10$ (see \textcolor{blue}{Appendix~E} for the case with ultra-strong coupling).
One can see similar scaling as in panel (a).

\section{Gravity-induced entanglement}
\begin{figure}[h]
\centering
\includegraphics[width=0.45\textwidth]{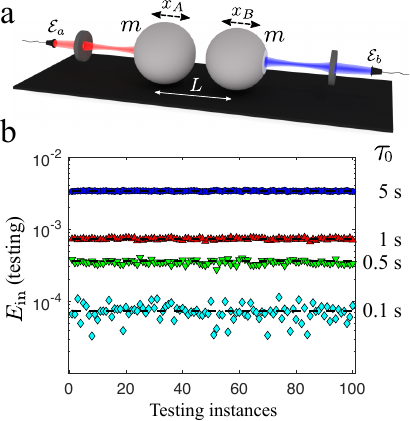}
\caption{Sensing gravity-induced entanglement (GIE) between masses.
(a) The setup of two trapped masses interacting gravitationally.
After a time $\tau_0$, two cavity modes are turned on, whose observables are processed to estimate the GIE.
(b) The estimated GIE for different initial evolution of the masses $\tau_0$ vs testing instances. One observes that the standard deviation $\delta_E<10^{-4}$.
The parameters used in simulations are $m=1$~kg made of Osmium with density $22.59$~g/cm$^3$, $\omega=0.1$~Hz with $\omega/\gamma \gg 1$, $r_0=1.73$, $\tau=1$~$\mu$s, $L_{a(b)}=25$~mm, laser wavelength $1064$~nm, $P_{a(b)}=50$~mW, $\{\kappa_{a},\kappa_b,\Delta_{a},\Delta_{b}\}=2.36\times 10^5$~Hz, $L \simeq 2R$ with $R$ the radius of the mass, and $\zeta/2=10^{-2}$.
See \textcolor{blue}{Appendix~G} for references of the parameters.}
\label{Fig4}
\end{figure}

We present an application of the entanglement sensing scheme to estimate gravity-induced entanglement (GIE) generated between masses.
Consider two identical spherical objects, each with mass $m$, trapped in a 1D harmonic potential. 
This configuration has been theoretically predicted to generate entanglement between the masses through gravitational interactions~\cite{krisnanda2020observable}.
Here, each mass is probed by a cavity mode, see Fig.~\ref{Fig4}(a).
The probes are turned on by the pump on the respective cavities, $\mathcal{E}_a$ (left) and $\mathcal{E}_b$ (right).
This way, the observables from the cavity modes can be processed through an output layer, which then produces an estimate of the GIE.

First, we consider the dynamics without the probes, in which the Hamiltonian reads
\begin{equation}
    H_0=\frac{\hbar \omega}{2}(\hat p_A^2+\hat x_A^2+\hat p_B^2+\hat x_B^2)-\frac{\hbar Gm}{\omega L^3}(\hat x_A-\hat x_B)^2,\label{EQ_H_2m}
\end{equation}
where $\hat x_{A(B)}$ denotes the dimensionless displacement of mass $A(B)$, $\omega$ the frequency of the trapping potentials, and $L$ the equilibrium distance between the masses. 
We have used $\hat x_{A(B)}= x_{A(B)}\sqrt{m\omega/\hbar}$ and $\hat p_{A(B)}=p_{A(B)}/\sqrt{\hbar m\omega}$, where $x_{A(B)}$ and $p_{A(B)}$ are the displacement and momentum operators, respectively.
The gravitational interaction is expanded from $-Gm^2/(L-(\hat x_A-\hat x_B)\sqrt{\hbar/m\omega})$ up to a quadratic term, $(\hat x_A-\hat x_B)^2$, which is necessary for entanglement generation as it contains non-local coupling $\propto \hat x_A\hat x_B$ acting on both masses.
We have neglected the constant and linear term $\propto(\hat x_A-\hat x_B)$ as the former is simply an energy offset and the latter a bi-local operator (cannot create entanglement) that constitutes to shifting the equilibrium position of the masses.
One can construct a set of LEs from Eq.~(\ref{EQ_H_2m}) with the addition of damping $\gamma$ and Brownian-like noises $\hat \xi_{A(B)}$ affecting the masses (see \textcolor{blue}{Appendix~G} for details).
As we deal with Gaussianity-preserving dynamics, we use the tools for CV systems.
This includes the description of the system within a CM and its evolution to $\bm V(\tau_0)$ from which properties of the system can be calculated (see \textcolor{blue}{Appendix~G}).

At time $\tau_0$, the probes are turned on, where the Hamiltonian (in a rotating frame with the frequency of the lasers) now reads
\begin{equation}
    \begin{aligned}
    H&=H_0+\hbar \Delta_{0a} \hat a^{\dagger}\hat a+\hbar \Delta_{0b} \hat b^{\dagger}\hat b+i\hbar \mathcal{E}_a(\hat a^{\dagger}-\hat a) \\
    &+i\hbar \mathcal{E}_b(\hat b^{\dagger}-\hat b)-\hbar G_{0a}\hat a^{\dagger}\hat a \hat x_A+\hbar G_{0b}\hat b^{\dagger}\hat b \hat x_B,\label{EQ_H_tot}
\end{aligned}
\end{equation}
where $\hat j = \hat a, \hat b$ denotes the annihilation operator of the left and right cavity mode, $\Delta_{0j}=\omega_{j}-\omega_{lj}$ the cavity-laser detuning, $\mathcal{E}_{j}=\sqrt{2P_j\kappa_j/\hbar \omega_{lj}}$ the driving strength of the cavity, $P_j$ the laser power with frequency $\omega_{lj}$, $\kappa_j=\pi c/2F_jL_j$ the cavity decay rate with finesse $F_j$ and length $L_j$, $G_{0j}=(\omega_j/L_j)\sqrt{\hbar/m \omega}$ the optomechanical coupling strength.
From Eq.~(\ref{EQ_H_tot}), one can construct a set of linearised LEs, which are then used to evolve the CM $\bm V(\tau_0)$ to $\bm V(\tau_0+\tau)$ at which the observables from the cavity modes are recorded.

In what follows, we take into account the features shown previously for entanglement sensing using generic systems. 
As the task is estimating entanglement of a two-mode CM (of the masses), at least 10 observables are required for recording.
This is taken from 10 independent CM elements of the joint cavity modes.
From the central limit theorem it follows that $\zeta/2 \propto 1/\sqrt{N_{\text {rep}}}$, where $N_{\text {rep}}$ is the number of repetitions that an element is measured. 
To make a comparison with entanglement measurement in Ref.~\cite{palomaki2013entangling} whereby $N_{\text {rep}}=10^4$, we shall assume error statistics with $\zeta \sim 2/\sqrt{N_{\text {rep}}} = 2\times 10^{-2}$.
As the initial CM at $t=0$, we use squeezed (local) thermal state for the masses $\mbox{diag}[e^{2r_0},e^{-2r_0},e^{2r_0},e^{-2r_0}](1+2\bar n)/2$ with $r_0$ being the squeezing strength and $\bar n$ the mean thermal phonon number, and vacuum for the cavity modes. 
The training is performed using random separable input states $\bm V_{\text{in}}(\tau_0)$, which are generated using random $\bar n>0$.
This is such that entanglement does not yet grow for initial thermal states within $\tau_0$.
On the other hand, testing is performed with $\bar n=0$.
For better precision, one can use time-multiplexing during the dynamics with the probes at $\{1,2,\cdots,\mathcal{T}\}\tau$. 

We present the estimated GIE for different initial accumulation time $\tau_0$ in Fig.~\ref{Fig4}(b).
We have taken $\mathcal{T}=4$ (see \textcolor{blue}{Appendix~H} for the scaling of standard deviation $\delta_E$ against $\mathcal{T}$), $N_{\text{tr}}=50$, and $N_{\text{te}}=100$.
The standard deviations of the GIE in Fig.~\ref{Fig4}(b) follow $\delta_E<10^{-4}$, which is two orders of magnitude better than the experimentally achieved $\sim 10^{-2}$ in Ref.~\cite{palomaki2013entangling}.
We also computed the estimated GIE from direct measurements, which is done by adding measurement errors directly to the elements of $\bm V(\tau_0)$.
In this case, standard deviation $\delta_E\sim 10^{-4}$ is only possible if the system permits three orders of magnitude weaker measurement error strength $\zeta/2 =10^{-5}$.
This demonstrates the efficiency of our method, which requires less number of single-shot measurements $N_{\text{rep}}$ to obtain precision comparable to measurements directly on the masses, i.e., with noisy $\bm V(\tau_0)$.

Figure~\ref{Fig4}(b) shows that our method is able to estimate GIE efficiently for $\tau_0=0.5$~s.
We note that this is shorter than the coherence times resulting from thermal photons from environment and collisions with air molecules (both in the range of about $5$~s) if the experiments were conducted on Earth with liquid Helium in ultrahigh vacuum~\cite{krisnanda2020observable}.


\section{DISCUSSION}
We have shown that a simple neural network (quantum reservoir processor) can be used for efficient estimation of quantum entanglement. Our main motivation for development of such a method is provided by present efforts to design experiments capable of detection of gravity-induced entanglement. 
The introduced method shows that the entanglement precision can be improved by two orders of magnitude from what was achieved in Ref.~\cite{palomaki2013entangling}.

The entanglement sensing step is crucial for masses initialised in natural Gaussian states
and any improvement on it relaxes other requirements of the setup.
The most direct one is the requirement on coherence times: since smaller values of entanglement become detectable, the system can be measured earlier.
With entanglement estimation accuracy on the order $10^{-4}$ detection of GIE could be performed within decoherence times available on Earth, whereas accuracy $10^{-2}$ would rather require an experiment in space~\cite{krisnanda2020observable}.
Moreover, in order to understand how other experimental parameters can be changed, let us recall that the \emph{figure of merit} for entanglement generated via gravity between trapped masses $m$ separated by a distance $L$ is given by $2 G m / \omega^2 L^3$, where $\omega$ characterises the trapping potential or spread of the initial wave function of each mass~\cite{krisnanda2020observable}.
Therefore, better entanglement precision also translates to smaller masses in the experiment that could be placed further apart.


The method presented in this paper also holds potential for other settings where one estimates entanglement of the easily accessed probes with precision advantage and reveals quantumness of a macroscopic mediating object.
In particular, this includes an extension of Refs.~\cite{lambert2013quantum,scholes2017using,collini2010coherently,panitchayangkoon2010long} towards showing quantum properties of photosynthetic bacteria~\cite{krisnanda2018probing} or that of a macroscopic mechanical membrane in the membrane-in-the-middle optomechanics setting~\cite{aspelmeyer2014cavity,paternostro2007creating,krisnanda2017revealing}.
Additionally, we note that our scheme can work not only for CV or discrete systems, but also hybrid configurations such as discrete systems as input and CV systems as the QN or vice versa (see \textcolor{blue}{Appendix~B}).

\begin{acknowledgements}
We thank Sanjib Ghosh, Kevin Dini, and Yvonne Gao for stimulating discussion.
T.K. and T.C.H.L. acknowledge the support by the Singapore Ministry of Education under its AcRF Tier 2 grant MOE2019-T2-1-004. 
T.P. is supported by the Polish National Agency for Academic Exchange NAWA Project No. PPN/PPO/2018/1/00007/U/00001. MP acknowledges the support by the European Union's Horizon 2020 FET-Open project  TEQ (766900), the Leverhulme Trust Research Project Grant UltraQuTe (grant RGP-2018-266), the Royal Society Wolfson Fellowship (RSWF/R3/183013), the UK EPSRC (EP/T028424/1), and the Department for the Economy Northern Ireland under the US-Ireland R\&D Partnership Programme (USI 175). 
\end{acknowledgements}

\vspace{0.1cm}
\noindent{\bf Author contributions:} TK, TP, and TCHL conceived the initial project direction; TK carried out all the calculations and derivations under the supervision of TP, MP, and TCHL; MP and TP assisted in designing the setup for gravity-induced entanglement; TK wrote the paper with contributions from TP, MP, and TCHL. All authors discussed the results and revised the paper.

\vspace{0.1cm}
\noindent{\bf Competing interests:} The authors declare no competing interests.

\vspace{0.1cm}
\noindent {\bf Data availability:} All data needed to evaluate the conclusions in the paper are present in the paper and/or the \textcolor{blue}{Appendix}. 

\clearpage

\appendix
\onecolumngrid
\section{Generic discrete systems: Entanglement estimation and scaling}\label{A_gds}
Here, we consider that all quantum systems, i.e., the input objects and the QN nodes, are qubits.
Each qubit has two energy levels, the ground state $|g\rangle$ and excited state $|e\rangle$.
Let us take the generic Hamiltonian in Eq.~(1) in the main text, where now $\hat a_n$ ($\hat b_m$) denotes the lowering operator $|g\rangle \langle e|$ for the $n$th input qubit ($m$th QN node).

In addition, the input and QN nodes may interact with their environment, adding an incoherent element to the dynamics. 
We consider a simple dissipative process such that the dynamics of the whole system is described within the Lindblad master equation
\begin{equation}\label{EQ_Dyn}
\dot \rho=-\frac{i}{\hbar}[H,\rho]+\sum_n \frac{\gamma_n}{2} \mathcal{L}(\rho,\hat a_n)+\sum_m \frac{\kappa_m}{2} \mathcal{L}(\rho,\hat b_m),
\end{equation}
where $\mathcal{L}(\rho,\hat X)\equiv 2\hat X \rho \hat X^{\dagger}-\{\hat X^{\dagger}\hat X,\rho \}$ and the QN nodes are initialised in their ground state $|g\rangle$.
The dissipation rate of the $n$th input and $m$th QN node are denoted by $\gamma_n$ and $\kappa_m$, respectively. These processes are not essential for our scheme, but are included to show robustness in their presence.
After a time $\tau$, the observables $\langle \hat O_{mk}\rangle=\mbox{tr}(\rho(\tau) \hat O_{mk})$ are recorded as a vector $\bm v$ and sent to a trained output layer (the training with ridge regression is described in the main text).
Note that the index $k$ denotes different observables from the same $m$th QN node.
The trained output layer is used to estimate the unique elements of the input state, giving us $\tilde \rho_{\text{in}}$, from which the entanglement is quantified using negativity $E=N_e(\tilde {\rho}_{\text{in}})$~\cite{negativity}.

In simulations, the system parameters are randomised in the same way as that described in Section~III in the main text.
The procedure to generate random input states for training $\{\rho_{\text{in},l}\}_{l=1}^{N_{\text{tr}}}$ and testing $\{\rho_{\text{in},l^{\prime}}\}_{l^{\prime}=1}^{N_{\text{te}}}$ are described below in Section~\ref{A_gris}.

\begin{figure}[h]
\centering
\includegraphics[width=0.6\textwidth]{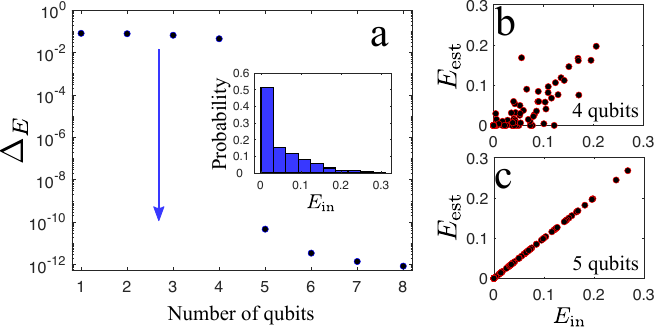}
\caption{(a) Estimation error vs number of qubits used in the QN.
The inset shows the profile of entanglement of the randomly generated input states in both training and testing. 
Panels (b) and (c) present explicit comparison between the estimated and input entanglement during testing where the QN is composed of $3$ and $4$ qubits, respectively.
}
\label{figs1_n}
\end{figure}

We tested the scheme to estimate entanglement of two-qubit input states.
The estimation error is plotted in Fig.~\ref{figs1_n}(a) against the number of qubits used in the QN. 
Here we recorded 3 observables from each qubit in the QN at $\tau=\pi/2\Gamma$, i.e., $\langle \sigma_x \rangle$, $\langle \sigma_y \rangle$, and $\langle \sigma_z \rangle$, where $\sigma_{\{x,y,z\}}$ stand for the Pauli matrices.  
Similar shift is seen where $\Delta_E\sim 10^{-11}$ is obtained for a QN with at least 5 qubits. 
This is because to fully characterise an $N$-qubit input state, one requires $2^{2N}-1$ parameters. 
This way, to estimate entanglement of a two-qubit input state ideally, at least 5 qubits are needed in the QN, corresponding to a total of 15 observables.
The direct comparison between the estimated and input entanglement can be seen in Figs.~\ref{figs1_n}(b) and (c) when the QN is composed of 4 and 5 qubits, respectively. 
We note that, in principle, if one were to record one observable from each qubit in the QN, it would require at least 15 qubits, which is too demanding to simulate on classical computers.
In this case, we show below that time-multiplexing is of help.

\begin{figure}[h]
\centering
\includegraphics[width=0.45\textwidth]{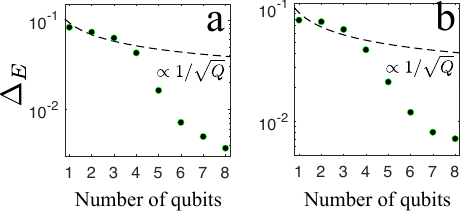}
\caption{(a) Estimation error plotted against the number of qubits $Q$ in QN.
Panel (b), taking only mean excitation from each QN node as the observable with time-multiplexing performed $\mathcal{T}=3$ times.
The scaling for SQL is given by the dashed curve in each panel.
}
\label{figs2_n}
\end{figure}

We also performed simulations by taking into account measurement errors with $\zeta=10^{-3}$.
We present the estimation errors in Fig.~\ref{figs2_n}(a) against the number of qubits used in the QN.
One can also utilise time-multiplexing with only measurements of $\langle \hat b^{\dagger}_m \hat b_m \rangle$ from each QN node.
An example of this is plotted in Fig.~\ref{figs2_n}(b), where the measurements are performed three times at $\tau=\{1,2,3\} \pi/2\Gamma$ on each QN node.
One can see that both panels in Fig.~\ref{figs2_n} show error scaling beyond the SQL (dashed curves) $\propto 1/\sqrt{\mathcal{N}}\propto 1/\sqrt{3Q}\propto 1/\sqrt{Q}$.

\section{Hybrid systems}
Here we show that the general scheme introduced in the main text is not limited to particular quantum systems (only CV or discrete systems). 
In what follows, we demonstrate this with a simple hybrid system: discrete input with CV QN.
It is important to note that as this involves discrete systems, the dynamics will not preserve Gaussianity of the CV systems in general. 
Covariance matrix does not fully describe the involved CV systems.
Therefore, we describe all quantum systems by their density matrices (truncated dimension for CV systems at $d = 20$).

Let us demonstrate sensing entanglement of two-qubit input with a single bosonic mode as the QN.
In particular, take the Hamiltonian as
\begin{equation}
    H=\hbar \sum_{n=1}^2 \Delta_n \hat a_n^{\dagger}\hat a_n + \hbar \Lambda \hat b^{\dagger}\hat b
    + \hbar \sum_{n=1}^2 K_n (\hat a_n \hat b^{\dagger} + \hat b \hat a_n^{\dagger}) + \hbar P(\hat b+ \hat b^{\dagger}),
\end{equation}
where $\hat a_n$ denotes the lowering operator ($|g\rangle \langle e|$) for the $n$th input qubit and $\hat b$ is the bosonic lowering operator for the single QN node.
For simplicity, we take the evolution as unitary, i.e., $\rho(t)=\hat U\rho(0)\hat U^{\dagger}$ with $\hat U=\exp(-iHt/\hbar)$.
The initial state is random $\rho_{\text{in}}$ for the qubits and vacuum $|0\rangle$ for the QN node.

The parameters are randomised as $\{\Delta_n,\Lambda,K_n,2P\}\in [1,2]\Gamma$.
As the readout, we take the mean excitation of the QN node $\langle \hat b^{\dagger} \hat b \rangle$ with time-multiplexing at $\mathcal{T}$ different times, i.e., $\tau=\{1,2,\cdots,\mathcal{T}\}\pi/10\Gamma$.
We tested this architecture for $10$ different realisations of the parameters, where in each we performed training with $N_{\text{tr}}=50$ and testing with $N_{\text{te}}=100$.
Our simulations show a transition to low entanglement estimation error $\Delta_E \sim 10^{-8}$ for $\mathcal{T}\ge 15$.
Again, this is because it requires at least 15 different parameters to characterise two-qubit input states.

Note that one can also apply the scheme to estimate entanglement of CV input systems (with truncated dimension) using discrete systems as the QN.
In general, the following requirements set the guidelines for a particular setup to be viable: 
\begin{enumerate}
\item The number of independent observables from the QN nodes has to be at least equal to the number of independent parameters required to characterise the state of the input objects. 
\item A dynamics ensuring that sufficient information about the inputs are carried forward to the QN observables. This requires the essential interactions between the input and QN nodes as well as within the QN nodes.
\item The independent QN observables may be obtained from different QN nodes or/and time-multiplexing (measurement of observables at different times). Normally, for simpler QN observables such as mean excitations, relatively richer dynamics is necessary. For CV systems, the latter can be achieved by, e.g., adding two-photon pumping for the QN nodes, having ultra-strong coupling between the involved quantum systems, or even nonlinearity. 
\end{enumerate}

\section{Generic CV systems: Details}\label{A_gcvsd}
From the Hamiltonian of Eq.~(1) in the main text, a set of LEs is obtained from the equations of motion in Heisenberg picture and the addition of noise terms:
\begin{eqnarray}
\dot{\hat a}_n&=&-(\gamma_n+i\Delta_n) \hat a_n-i\sum_m K_{nm}\hat b_m +\sqrt{2\gamma_n}\;\hat a_n^{\text{in}}, \nonumber \\
\dot{\hat b}_m&=&-(\kappa_m+i\Lambda_m) \hat b_m -i\sum_n K_{nm} \hat a_n -i\sum_{m^{\prime}}J_{mm^{\prime}}\hat b_{m^{\prime}}-iP_m+ \sqrt{2\kappa_m}\; \hat b_m^{\text{in}},
\end{eqnarray}
where $\hat a_n^{\text{in}}$ and $\hat b_m^{\text{in}}$ are zero mean Gaussian noise operators with correlation functions $\langle \hat a_n^{\text{in}}(t) \hat a_{n^{\prime}}^{\text{in},\dagger}(t^{\prime})\rangle=\delta_{nn^{\prime}}\delta(t-t^{\prime})$ and $\langle \hat b_m^{\text{in}}(t) \hat b_{m^{\prime}}^{\text{in},\dagger}(t^{\prime})\rangle=\delta_{mm^{\prime}}\delta(t-t^{\prime})$~\cite{walls2007quantum}.
This allows us to write the LEs in terms of dimensionless position and momentum quadratures
\begin{eqnarray}
\dot{\hat q}_n&=&-\gamma_n\hat q_n+\Delta_n \hat r_n+\sum_m K_{nm}\hat p_m +\sqrt{2\gamma_n}\;\hat q_n^{\text{in}} \nonumber \\
\dot{\hat r}_n&=&-\gamma_n\hat r_n-\Delta_n \hat q_n-\sum_m K_{nm}\hat x_m +\sqrt{2\gamma_n}\;\hat r_n^{\text{in}} \nonumber \\
\dot{\hat x}_m&=&-\kappa_m\hat x_m+\Lambda_m \hat p_m+\sum_n K_{nm}\hat r_n+\sum_{m^{\prime}} J_{mm^{\prime}}\hat p_{m^{\prime}} +\sqrt{2\kappa_m}\;\hat x_m^{\text{in}} \nonumber \\
\dot{\hat p}_m&=&-\kappa_m\hat p_m-\Lambda_m \hat x_m-\sum_n K_{nm}\hat q_n-\sum_{m^{\prime}} J_{mm^{\prime}}\hat x_{m^{\prime}} -\sqrt{2}P_m+\sqrt{2\kappa_m}\;\hat p_m^{\text{in}}.\label{EQ_LV_GS}
\end{eqnarray}
We have used the following quadrature relations:
\begin{eqnarray}
\hat q_n&=&\frac{\hat a_n+\hat a_n^{\dagger}}{\sqrt{2}}, \:\:\hat r_n=\frac{\hat a_n-\hat a_n^{\dagger}}{i\sqrt{2}},\:\: 
\hat x_m=\frac{\hat b_m+\hat b_m^{\dagger}}{\sqrt{2}}, \:\:\hat p_m=\frac{\hat b_m-\hat b_m^{\dagger}}{i\sqrt{2}},\nonumber \\
\hat q_n^{\text{in}}&=&\frac{\hat a_n^{\text{in}}+(\hat a_n^{\text{in}})^{\dagger}}{\sqrt{2}}, \:\:\hat r_n^{\text{in}}=\frac{\hat a_n^{\text{in}}-(\hat a_n^{\text{in}})^{\dagger}}{i\sqrt{2}}\:\: 
\hat x_m^{\text{in}}=\frac{\hat b_m^{\text{in}}+(\hat b_m^{\text{in}})^{\dagger}}{\sqrt{2}}, \:\:\hat p_m^{\text{in}}=\frac{\hat b_m^{\text{in}}-(\hat b_m^{\text{in}})^{\dagger}}{i\sqrt{2}}.
\end{eqnarray}
The LEs in Eq.~(\ref{EQ_LV_GS}) can be written in a matrix form 
\begin{equation}
    \dot {\bm{u}}(t) = {\bm A} {\bm{u}}(t) + {\bm{h}}(t),\label{EQ_mLGV_GS}
\end{equation}
where the vector
\begin{eqnarray}
    \bm{u}=[\hat q_1,\hat r_1,\hat q_2,\hat r_2,\hat x_1,\hat p_1,\hat x_2,\hat p_2,\cdots,\hat x_M,\hat p_M]^T
\end{eqnarray}
contains the quadratures, the drift matrix reads
\begin{equation}
    \bm{A}=\left[\begin{array}{ccccccccccc} -\gamma_1 & \Delta_1 & 0 & 0 & 0&K_{11}&0&K_{12}&\cdots & 0&K_{1M} \\
    -\Delta_1 & -\gamma_1 & 0 & 0 & -K_{11}&0&-K_{12}&0&\cdots &-K_{1M}&0\\
    0 & 0 & -\gamma_2 & \Delta_2 & 0&K_{21}&0&K_{22}&\cdots &0&K_{2M}\\
    0 & 0 & -\Delta_2 & -\gamma_2& -K_{21}&0&-K_{22}&0&\cdots &-K_{2M}&0 \\
    0 & K_{11} & 0 & K_{21} & -\kappa_1&\Lambda_1&0&J_{11}&\cdots &0&J_{1M}\\
    -K_{11}&0&-K_{21}&0&-\Lambda_1&-\kappa_1&-J_{11}&0&\cdots & -J_{1M}&0\\
    0&K_{12}&0&K_{22}&0&J_{11}&-\kappa_2&\Lambda_2&\cdots & 0 &J_{2M}\\
    -K_{12}&0&-K_{22}&0&-J_{11}&0&-\Lambda_2&-\kappa_2&\cdots &-J_{2M}&0\\
    \vdots&\vdots&\vdots&\vdots&\vdots&\vdots&\vdots&\vdots&\ddots&\vdots&\vdots\\
    0&K_{1M}&0&K_{2M}&0&J_{1M}&0&J_{2M}&\cdots&-\kappa_M&\Lambda_M\\
    -K_{1M}&0&-K_{2M}&0&-J_{1M}&0&-J_{2M}&0&\cdots&-\Lambda_M&-\kappa_M \end{array}\right],
\end{equation}
and the vector 
\begin{equation}
    \bm{h}=\left[\begin{array}{c}0\\
    0\\
    0\\
    0\\
    0\\
    -\sqrt{2}P_1\\
    0\\
    -\sqrt{2}P_2\\
    \vdots\\
    0\\
   -\sqrt{2}P_M\end{array}\right]+\left[\begin{array}{c}\sqrt{2\gamma_1}\hat q_1^{\text{in}}\\
    \sqrt{2\gamma_1}\hat r_1^{\text{in}}\\
    \sqrt{2\gamma_2}\hat q_2^{\text{in}}\\
    \sqrt{2\gamma_2}\hat r_2^{\text{in}}\\
    \sqrt{2\kappa_1}\hat x_1^{\text{in}}\\
    \sqrt{2\kappa_1}\hat p_1^{\text{in}}\\
    \sqrt{2\kappa_2}\hat x_2^{\text{in}}\\
    \sqrt{2\kappa_2}\hat p_2^{\text{in}}\\
    \vdots\\
    \sqrt{2\kappa_M}\hat x_M^{\text{in}}\\
    \sqrt{2\kappa_M}\hat p_M^{\text{in}}\end{array}\right]
\end{equation}
contains the pump and noise terms. The solution to Eq.~(\ref{EQ_mLGV_GS}) is given by \begin{equation}
    {\bm u}(t)={\bm W}_+(t){\bm u}(0)+{\bm W}_+(t)\int_0^t dt^{\prime} {\bm W}_-(t^{\prime}){\bm h}(t^{\prime}),\label{EQ_gomu}
\end{equation}
where ${\bm W}_{\pm}(t)=\exp{(\pm {\bm A}t)}$.
One can then form the CM at time $t$ and show that it follows
\begin{equation}
    \frac{d{\bm V}}{dt}={\bm A}{\bm V}(t)+{\bm V}(t){\bm A}^T+{\bm D},\label{EQ_nomi}
\end{equation}
where ${\bm D}=\mbox{diag}[\gamma_1,\gamma_1,\gamma_2,\gamma_2,\kappa_1,\kappa_1,\kappa_2,\kappa_2,\cdots,\kappa_M,\kappa_M]$.

In simulations, the initial CM of the system is taken as
\begin{equation}
    \bm{V}(0)=\left[\begin{array}{cc}\bm{V}_{\text{in}}(0)&\bm{0}\\
    \bm{0}&\bm{V}_{\text{qn}}(0)\end{array} \right],
\end{equation}
where $\bm{V}_{\text{in}}(0)$ is the initial CM for the input objects, whereas that for the QN nodes is initiated with vacuum $\bm{V}_{\text{qn}}(0)=\openone/2$.

\section{The generation of random input states for generic dynamics}\label{A_gris}

For CV systems, the random two-mode input CMs are generated dynamically as follows.
We take a Hamiltonian of the form
\begin{equation}
    H=\hbar \Delta_1 \hat a_1^{\dagger}\hat a_1 +\hbar \Delta_2 \hat a_2^{\dagger}\hat a_2 
    + \hbar K (\hat a_1 \hat a_2^{\dagger} + \hat a_2 \hat a_1^{\dagger}) + \hbar P^{\prime}_1(\hat a_1 \hat a_1 + \hat a_1^{\dagger} \hat a_1^{\dagger}) + \hbar P^{\prime}_2(\hat a_2 \hat a_2 + \hat a_2^{\dagger} \hat a_2^{\dagger}),
\end{equation}
where the two modes are coupled and pumped (two-photon drives).
The drives have terms similar to single mode squeezing operations. 
As the initial CM, we take vacuum $\openone/2$.
The parameters are taken as random, i.e., $(\Delta_1,\Delta_2,K,P^{\prime}_1,P^{\prime}_1)\in (1,1,1,0.3,0.3) [0,1]\Gamma$, and the evolution time $\tau_0=\pi /2\Gamma$.
The random input CMs used for CV systems in the main text are sampled from $\bm V(\tau_0)$.
The entanglement profile resulting from this distribution is plotted in the inset of Fig.~2(a) in the main text.
For the scheme where learning is done with non-entangled states, the evolution time is taken as $\tau_0=1 /2\Gamma$ with thermal states (CM is $3\openone/2$) as the initial condition.

For discrete systems (qubits), the random input states are sampled as follows.
\begin{eqnarray}
Z/10 &=& 2(\upsilon_1 + i\upsilon_2) - (1+i)\mathcal{J}+ h.c.,\nonumber \\
\rho_{\text{in}}&=&ZZ^{\dagger}/\text{tr}(ZZ^{\dagger}),
\end{eqnarray}
where $\upsilon_{1,2}$ is a random $4\times 4$ matrix whose elements are sampled from standard normal distribution and $\mathcal{J}$ is a $4\times 4$ matrix of ones.
This sampling results in entanglement profile shown in the inset of Fig.~\ref{figs1_n}(a).
For generating classically correlated states, one can simply destroy the entanglement in $\rho_{\text{in}}$ by projective measurements on one input object, i.e., $\rho_{\text{in, sep}}=\Pi_1\rho_{\text{in}}\Pi_1+\Pi_2\rho_{\text{in}}\Pi_2$ with $\{\Pi_1,\Pi_2\}$ as random projection operators and $\Pi_1+\Pi_2=\openone$.

\section{Entanglement estimation for CV systems using mean excitations}\label{A_eecvme}

\begin{figure*}[!h]
\centering
\includegraphics[width=0.8\textwidth]{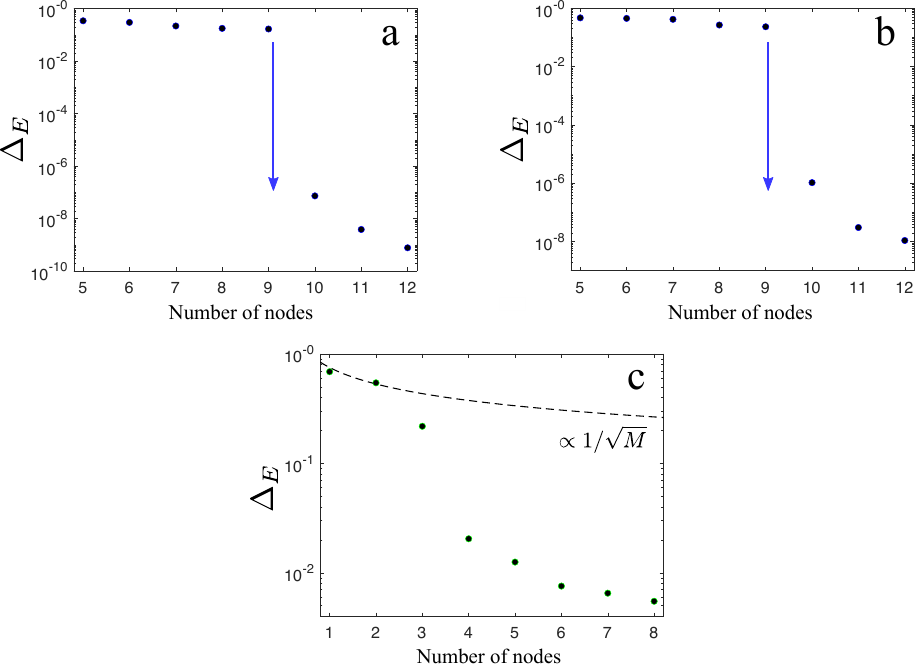}
\caption{Performance of entanglement sensing for CV systems using mean excitations $\{\langle \hat b_m^{\dagger}\hat b_m \rangle\}$: (a) with the addition of two-photon pump or (b) the use of ultra-strong coupling.
Measurement errors are not taken into account in panels (a) and (b).
(c) The estimation error as in panel (b) with time-multiplexing at $\tau=\{1,2,3\}\pi/2\Gamma$, taking into account measurement errors with $\zeta=10^{-3}$.
The dashed curve represents the SQL scaling.
}
\label{figs3}
\end{figure*}

In the main text, we have demonstrated that by taking 3 observables from each QN node, the shift to low estimation error requires at least 4 QN nodes, see Fig.~2(a).
Here, we present the case where we utilise the mean excitation from each QN node instead.
In this case, we present two options both of which require additional (necessary) ingredient. 
First, one can add two-photon pump to each QN node (the two-photon pump can be relatively weaker in strength than the single-photon pump). 
This is carried out by adding $\sum_m P_m^{\prime}(\hat b_m\hat b_m+\hat b_m^{\dagger}\hat b_m^{\dagger})$ to the Hamiltonian of Eq.~(1) in the main text.
For simulations, we take $P_m^{\prime}\in [0,1]\Gamma/10$.
We present the estimation error in Fig.~\ref{figs3}(a), where the shift to low estimation error is achieved for a QN having at least 10 nodes.
For option two, the interactions between the QN nodes are taken following the ultra-strong coupling type. 
In this case, one simply replaces the operator function in Eq.~(1) with $\mathcal{F}(\hat b_m,\hat b_{m^{\prime}})=(\hat b_m+\hat b_m^{\dagger})(\hat b_{m^{\prime}} +\hat b_{m^{\prime}}^{\dagger})$.
Similarly, the estimation error is plotted in Fig.~\ref{figs3}(b).

Similar to Fig.~3(b) in the main text, we present the estimation error in Fig.~\ref{figs3}(c) for CV systems using ultra-strong coupling.
The time-multiplexing is $\mathcal{T}=3$ and the strength of measurement errors is $\zeta=10^{-3}$.

We appreciate that the function of the quantum reservoir is to map the input state (or its parameters) to the measured local observables. 
The mapping is unknown, as the parameters defining the reservoir are allowed to be random. 
However, one can assume that the measured local observables attained by the mapping must contain sufficient information to be a representation of the initial input state. 
Apparently, without two-photon pumping or ultra-strong coupling information is not well spread, i.e., many distinct states are mapped to the same QN state, and therefore, information is lost. 
The two-photon pumping creates and destroy two excitations on QN nodes and ultra-strong coupling ensures creation/annihilation of two excitations in addition to the normal hopping type coupling between the QN nodes.
Both processes result in more states being populated, explore higher-dimensional subspace of the QN Hilbert space, which makes the spread of information more complex.

\section{Learning with non-entangled states: Results}\label{A_lnes}
\begin{figure}[h]
\centering
\includegraphics[width=0.4\textwidth]{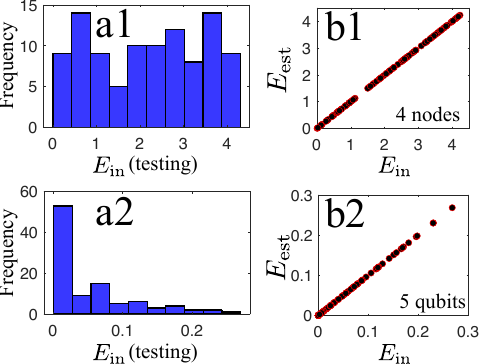}
\caption{Learning with non-entangled states.
For CV systems: (a1) entanglement profile of the input states in testing and (b1) estimated vs input entanglement in testing using 4 QN nodes.
The corresponding plots for discrete systems are presented in panels (a2) and (b2) using 5 qubits in the QN.
All input states for training have zero entanglement $E_{\text{in}}=0$.
}
\label{figs4}
\end{figure}

As described in the main text (Fig.~2) and Section~\ref{A_gds} above, entanglement sensing may be performed using generic CV or discrete systems.
There, the training and testing both use random entangled input states, with entanglement profile given in the inset of Fig.~2(a) and Fig.~\ref{figs1_n}(a).
We present similar analysis in Fig.~\ref{figs4} where the training only utilises separable input states.
For CV systems, panel (a1) shows the entanglement profile of the input states used in testing and (b1) the comparison between estimated and input entanglement for a QN composed of 4 nodes.
Similarly, the case for discrete systems are presented in panels (a2) and (b2) using a QN composed of 5 qubits.
One can see that panels (b1) and (b2) are similar to Fig.~2(c) in the main text and Fig.~\ref{figs1_n}(c), respectively.

\section{Gravity-induced entanglement: Details}\label{A_gied}
For the case without probes, the equations of motion in Heisenberg picture read
\begin{eqnarray}
\dot {\hat x}_A&=&\omega \hat p_A, \:\: \dot {\hat x}_B=\omega \hat p_B,\nonumber \\
\dot {\hat p}_A&=&-\omega(1-\eta)\hat x_A-\omega \eta \hat x_B -\gamma \hat p_A +\hat \xi_A,\nonumber \\
\dot {\hat p}_B&=&-\omega(1-\eta)\hat x_B-\omega \eta \hat x_A -\gamma \hat p_B +\hat \xi_B,\label{EQ_LGV_G_1}
\end{eqnarray}
where $\eta\equiv 2Gm/\omega^2L^3$, $\gamma$ represents the damping of each mass, and $\hat \xi_{A(B)}$ the Brownian-like noise for the masses. 
We assume high mechanical quality factor, $\omega/\gamma \gg 1$, where the noises can be treated as uncoloured noise with correlation function $\langle \hat \xi_j(t)\hat \xi_j(t^{\prime})+\hat \xi_j(t^{\prime})\hat \xi_j(t)\rangle/2=\gamma (2\bar n+1)\delta(t-t^{\prime})$, $j=\{A,B\}$~\cite{giovannetti2001phase,benguria1981quantum}.
The thermal phonon number $\bar n$ is related to the temperature of the environment $T$ as $\bar n=1/(\exp(\hbar \omega/k_BT)-1)$.

The matrix form of the LEs in (\ref{EQ_LGV_G_1}) is written as $ \dot {\bm{u}}(t) = {\bm A} {\bm{u}}(t) + {\bm{h}}(t)$, where
\begin{equation}
    \bm{u}=[\hat x_A,\hat p_A,\hat x_B,\hat p_B]^T,
\end{equation}
\begin{equation}
    \bm{A}=\left[ \begin{array}{cccc}0&\omega&0&0\\
    -\omega(1-\eta)&-\gamma&-\omega \eta &0\\
    0&0&0&\omega\\
    -\omega \eta&0&-\omega(1-\eta)&-\gamma \end{array}\right],
\end{equation}
and
\begin{equation}
    \bm{h}=[0,\hat \xi_A,0,\hat \xi_B]^T.
\end{equation}
The solution to the quadratures $\bm{u}(t)$ and CM $\bm{V}(t)$ are obtained as in Eqs.~(\ref{EQ_gomu}) and (\ref{EQ_nomi}), where $\bm{D}=\mbox{diag}[0,\gamma(2\bar n+1),0,\gamma(2\bar n+1)]$.

As the initial state for each mass, we take thermal squeezed state, i.e., $\bm{V}(0)=\mbox{diag}[e^{2r_0},e^{-2r_0},e^{2r_0},e^{-2r_0}](1+2\bar n)/2$.
The system is evolved for a time $\tau_0$, after which one obtains $\bm{u}(\tau_0)$ and $\bm{V}(\tau_0)$.

When the probes are turned on at $\tau_0$, the new dynamics is now described as follows.
From the Hamiltonian of Eq.~(5) in the main text, the new LEs read
\begin{eqnarray}
    \dot {\hat x}_A&=&\omega \hat p_A, \:\: \dot {\hat x}_B=\omega \hat p_B,\nonumber \\
    \dot {\hat p}_A&=&-\omega(1-\eta)\hat x_A-\omega \eta \hat x_B -\gamma \hat p_A +G_{0a}\hat a^{\dagger} \hat a+\hat \xi_A,\nonumber \\
\dot {\hat p}_B&=&-\omega(1-\eta)\hat x_B-\omega \eta \hat x_A -\gamma \hat p_B -G_{0b}\hat b^{\dagger} \hat b+\hat \xi_B,\nonumber \\
    \dot {\hat a}&=&-(\kappa_a+i\Delta_{0a})\hat a+iG_{0a}\hat a \hat x_A+\mathcal{E}_a+\sqrt{2\kappa_a}\;\hat a^{\text{in}} \nonumber \\
    \dot {\hat b}&=&-(\kappa_b+i\Delta_{0b})\hat b-iG_{0b}\hat b \hat x_B+\mathcal{E}_b+\sqrt{2\kappa_b}\;\hat b^{\text{in}},
\end{eqnarray}
where $\hat a^{\text{in}}$ and $\hat b^{\text{in}}$ are Gaussian noise operators with $\langle \hat a^{\text{in}}(t)\hat a^{\text{in},\dagger}(t^{\prime})\rangle=\delta(t-t^{\prime})$ and $\langle \hat b^{\text{in}}(t)\hat b^{\text{in},\dagger}(t^{\prime})\rangle=\delta(t-t^{\prime})$~\cite{walls2007quantum}.

The linearised version of the LEs is obtained through the following  transformations
\begin{equation*}
    \begin{aligned}
    \hat x_A&\rightarrow x_{As}+\delta \hat x_A,\:\hat p_A\rightarrow p_{As}+\delta \hat p_A,  \\
    \hat x_B&\rightarrow x_{Bs}+\delta \hat x_B,\:\hat p_B\rightarrow p_{Bs}+\delta \hat p_B,  \\
    \hat a&\rightarrow\alpha_s+\delta \hat a, \\
    \hat b&\rightarrow\beta_s+\delta \hat b,
\end{aligned}
\end{equation*}
and by ignoring any nonlinear term such as $\delta \hat a^{\dagger}\delta \hat a$, $\delta \hat a \delta \hat x_A$, $\delta \hat b^{\dagger}\delta \hat b$, and $\delta \hat b \delta \hat x_B$ in  the fluctuation operators.
In what follows, we will consider a much shorter evolution time such that one may neglect the contribution from the gravitational coupling ($\eta \approx 0$).
In particular, we have
\begin{eqnarray}
    \delta \dot {\hat x}_{A}&=&\omega \delta \hat p_{A}, \:\: \delta \dot {\hat x}_B=\omega \delta \hat p_B,\nonumber \\
    \delta \dot {\hat p}_A&=&-\omega \delta \hat x_A-\gamma \delta \hat p_A +G_{a}\delta \hat x_a+\hat \xi_A,\nonumber \\
\delta \dot {\hat p}_B&=&-\omega \delta \hat x_B-\gamma \delta \hat p_B -G_{b}\delta \hat x_b+\hat \xi_B,\nonumber \\
    \delta \dot {\hat x}_a &=&-\kappa_a \delta \hat x_a+\Delta_a \delta \hat p_a +\sqrt{2\kappa_a}\;\hat x_a^{\text{in}},\nonumber \\
    \delta \dot {\hat p}_a&=&-\kappa_a \delta \hat p_a-\Delta_a \delta \hat x_a +G_a \delta \hat x_A+\sqrt{2\kappa_a} \;\hat p_a^{\text{in}}\nonumber \\
    \delta \dot {\hat x}_b &=&-\kappa_b \delta \hat x_b+\Delta_b \delta \hat p_b +\sqrt{2\kappa_b}\;\hat x_b^{\text{in}},\nonumber \\
    \delta \dot {\hat p}_b&=&-\kappa_b \delta \hat p_b-\Delta_b \delta \hat x_b -G_b \delta \hat x_B+\sqrt{2\kappa_b} \;\hat p_b^{\text{in}},\label{EQ_LGV_final}
\end{eqnarray}
where
\begin{eqnarray}
    p_{As}&=&0, \:\: p_{Bs}=0,\nonumber \\
    x_{As}&=&\frac{G_{0a}|\alpha_s|^2}{\omega}, \:\: x_{Bs}=-\frac{G_{0b}|\beta_s|^2}{\omega}, \nonumber \\
    \alpha_s&=&\frac{|\mathcal{E}_a|}{\sqrt{\kappa_a^2+\Delta_a^2}},\:\: \beta_s=\frac{|\mathcal{E}_b|}{\sqrt{\kappa_b^2+\Delta_b^2}}.
\end{eqnarray}
In these equations, we have introduced the quantities $\Delta_a=\Delta_{0a}-G_{0a}x_{As}$, $\Delta_b=\Delta_{0b}+G_{0b}x_{Bs}$, $G_a=G_{0a}\alpha_s\sqrt{2}$, and $G_b=G_{0b}\beta_s\sqrt{2}$.
Note that $\alpha_s$ and $\beta_s$ have been assumed real, which can be done by tuning the phase of the laser $\mathcal{E}_a$ and $\mathcal{E}_b$, respectively.
We have also used the following quadrature relations:
\begin{eqnarray}
\hat x_a&=&\frac{\hat a+\hat a^{\dagger}}{\sqrt{2}}, \:\:\hat p_a=\frac{\hat a-\hat a^{\dagger}}{i\sqrt{2}},\:\:
\hat x_b=\frac{\hat b+\hat b^{\dagger}}{\sqrt{2}}, \:\:\hat p_b=\frac{\hat b-\hat b^{\dagger}}{i\sqrt{2}},\nonumber \\
\hat x_a^{\text{in}}&=&\frac{\hat a^{\text{in}}+(\hat a^{\text{in}})^{\dagger}}{\sqrt{2}}, \:\:\hat p_a^{\text{in}}=\frac{\hat a^{\text{in}}-(\hat a^{\text{in}})^{\dagger}}{i\sqrt{2}}\:\:
\hat x_b^{\text{in}}=\frac{\hat b^{\text{in}}+(\hat b^{\text{in}})^{\dagger}}{\sqrt{2}}, \:\:\hat p_b^{\text{in}}=\frac{\hat b^{\text{in}}-(\hat b^{\text{in}})^{\dagger}}{i\sqrt{2}}.
\end{eqnarray}

One can write the LEs for the fluctuation of the quadratures in (\ref{EQ_LGV_final}) as $\dot {\bm{u}}(t) = {\bm A} {\bm{u}}(t) + {\bm{h}}(t)$, where now 
\begin{equation}
    \bm{u}=[\delta \hat x_A,\delta \hat p_A,\delta \hat x_B,\delta \hat p_B,\delta \hat x_a,\delta \hat p_a,\delta \hat x_b,\delta \hat p_b]^T,
\end{equation}
\begin{equation}
    \bm{A}=\left[ \begin{array}{cccccccc}0&\omega&0&0&0&0&0&0\\
    -\omega&-\gamma&0&0&G_a&0&0&0\\
    0&0&0&\omega&0&0&0&0\\
    0&0&-\omega&-\gamma&0&0&-G_b&0\\
    0&0&0&0&-\kappa_a&\Delta_a&0&0\\
    G_a&0&0&0&-\Delta_a&-\kappa_a&0&0\\
    0&0&0&0&0&0&-\kappa_b&\Delta_b\\
    0&0&-G_b&0&0&0&-\Delta_b&-\kappa_b\\
    \end{array}\right],
\end{equation}
and
\begin{equation}
    \bm{h}=[0,\hat \xi_A,0,\hat \xi_B,\hat x_a^{\text{in}},\hat p_a^{\text{in}},\hat x_b^{\text{in}},\hat p_b^{\text{in}}]^T.
\end{equation}
The solution $\bm{u}(t)$ and $\bm{V}(t)$ are obtained from Eqs.~(\ref{EQ_gomu}) and (\ref{EQ_nomi}), where $\bm{D}=\mbox{diag}[0,\gamma(2\bar n+1),0,\gamma(2\bar n+1),\kappa_a,\kappa_a,\kappa_b,\kappa_b]$.
As the initial CM, we take $\bm{V}(\tau_0)$ for the masses and vacuum $\openone/2$ for the cavity modes.

The parameters chosen in the main text (see the caption of Fig.~4) are motivated as follows.
Mechanical mirrors of mass $m\sim 1$~kg and frequency $\omega\sim 0.1$~Hz have been cooled down near their ground state~\cite{abbott2009observation}, see also Ref.~\cite{whittle2021approaching}.
The initial state for each mass (squeezed thermal) can be prepared by appropriate optical driving~\cite{vanner2013cooling,rashid2016experimental}, and the strength $r_0=1.73$ is motivated by the squeezing of light mode~\cite{vahlbruch2016detection} and advances in the state transfer in optomechanics~\cite{aspelmeyer2014cavity}.
Cavity length ($25$~mm) and laser wavelength $1064$~nm are typical in optomechanics~\cite{groblacher2009observation}, see also Ref.~\cite{aspelmeyer2014cavity} for cavity finesse up to $10^5$.
For example, a cavity finesse $F=8\times 10^{4}$ gives a decay rate $\Gamma=\pi c/2FL_{a(b)}\approx 2.36\times 10^{5}$~Hz.
This is used in simulations as a basis for the random cavity decay rates and effective detunings, i.e., $\{\kappa_{a},\kappa_b,\Delta_{a},\Delta_{b}\}\in[1,2]\Gamma.$

\section{Gravity-induced entanglement: Error scaling}\label{A_giees}

\begin{figure}[h]
\centering
\includegraphics[width=0.4\textwidth]{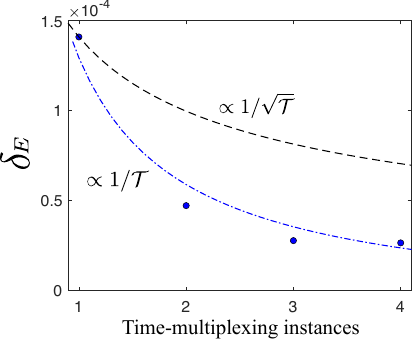}
\caption{Error scaling. The standard deviation for the setup in Fig.~4(b) in the main text for $\tau_0=5$~s vs instances of time-multiplexing $\mathcal{T}$.
}
\label{figs5}
\end{figure}

Figure~\ref{figs5} presents the entanglement estimation error against time-multiplexing instances $\mathcal{T}$ for the case of Fig.~4(b) with $\tau_0=5$~s.
The estimation error is taken as standard deviation, 
\begin{equation}
\delta_E=\sqrt{\sum_{l^{\prime}}^{N_{\text{te}}} \frac{(E_{\text{est},l^{\prime}}-E_{\text{in}})^2}{N_{\text{te}}-1}}.
\end{equation}
One can see that the scaling of the estimation error is beyond the SQL (dashed curve).
In fact, it follows a Heisenberg-like scaling with $\delta_E \propto \mathcal{T}^{-1}$ (dashed-dotted curve).
We note that for $\mathcal{T}=4$, the estimation error is comfortably below $10^{-4}$, which is two orders of magnitude lower than what was experimentally achieved in Ref.~\cite{palomaki2013entangling}. 

\bibliography{Bibliography}

\end{document}